\documentclass{moriond}

\usepackage{amsmath}
\usepackage{amssymb}
\usepackage{slashed}

\def\be{\begin{equation}}
\def\ee{\end{equation}}
\def\bea{\begin{eqnarray}}
\def\eea{\end{eqnarray}}

\begin{document}
\vspace*{4cm}
\title{Exploring $t$-Channel Models for Dark Matter}

\author{David Cabo-Almeida}
\address{
Dipartimento di Scienze Matematiche e Informatiche, Scienze Fisiche e Scienze della Terra\\
Universita degli Studi di Messina, 
Via Ferdinando Stagno d'Alcontres 31, I-98166 Messina, Italy\\[2mm]
INFN Sezione di Catania Via Santa Sofia 64, I-95123 Catania, Italy\\[2mm]
Departament de Física Quàntica i Astrofísica, Universitat de Barcelona,\\
Martí i Franquès 1, E08028 Barcelona, Catalunya, Spain\\[2mm]
Institut de Ciències del Cosmos (ICCUB), Universitat de Barcelona,\\
Martí i Franquès 1, E08028 Barcelona, Catalunya, Spain}

\maketitle\abstracts{
We report on a comprehensive study of the Direct Detection phenomenology of singlet Dark Matter $t$-channel portal models. For that purpose, we present a complete computation of the loop-induced direct detection cross-section for both scalar and fermionic Dark Matter candidates. We complete the study by comparing the results with current and future bounds from Direct Detection experiments and requiring the correct Dark Matter relic density.
}
\section{Introduction}
Introducing a cosmologically-stable particle into the Standard Model (SM) stands as one of the most favored approaches to unravelling the Dark Matter (DM) enigma. Among the variety of proposed particle physics frameworks, Weakly Interacting Massive Particles (WIMPs) have seized the spotlight due to their testability. Recent years have witnessed significant strides in Dark Matter Direct Detection (DD) experiments, probing numerous particle physics models. This prompts an intriguing question: Can current experimental sensitivity, or that anticipated in the next generation of detectors, effectively scrutinize models where interactions pertinent to Direct Detection emerge at the loop level?

One noteworthy example of such scenarios lies within the class of DM simplified models termed ``$t$-channel portals''. Here, DM exhibits a Yukawa-like interaction with an SM fermion and a novel particle possessing nontrivial quantum characteristics under the SM gauge group. The interactions of these new particles are governed by gauge invariance and determined by their representations within the SM gauge group.

This talk delves into the direct detection prospects within $t$-channel portals. We present the most current assessment of the DM scattering cross-section and scattering rate over nucleons for several variants of this mode. Considering scalar (real and complex) and fermionic (Dirac and Majorana) DM candidates we compare the theoretical prediction with the constraints from Direct Detection experiments as well as the requisite for the correct DM relic density, under the assumption of freeze-out paradigm.
\section{Model and Direct Detection}
The most generic way to write this kind of $t$-channel portal models for the scalar/fermionic scenario are described by the following interaction:
\bea
\nonumber
\mathcal{L}_{\text{scalar/fermi}}&=& \Gamma_L^{f_i} \bar{f}_i P_R \Psi_{f_i/\mathrm{DM}} \Phi_{\mathrm{DM}/f_i}+\Gamma_R^{f_i} \bar{f}_i P_L \Psi_{f_i/\mathrm{DM}} \Phi_{\mathrm{DM}/f_i}+\text { h.c. } \\
& &+\lambda_{1 H \Phi}\left(\Phi_{\mathrm{DM}/f_i}^{\dagger} \Phi_{\mathrm{DM}/f_i}\right)\left(H^{\dagger} H\right)+\lambda_{2 H \Phi}\left(\Phi_{\mathrm{DM}/f_i}^{\dagger} T_{\Phi}^a \Phi_{\mathrm{DM}/f_i}\right)\left(H^{\dagger} \frac{\sigma^a}{2} H\right),
\label{eq:model}
\eea
where the DM field has been label as $\Phi_{\mathrm{DM}} (\Psi_{\mathrm{DM}})$ while $\Psi_{f_i} (\Phi_{f_i})$ represent the mediator. To maintain the notation as general as possible, we have introduced generic couplings $\Gamma_L^{f_i},\Gamma_R^{f_i}$ of the new fermions with both left-handed and right-handed SM fermions.
Moreover, the Lagrangian is formulated under the assumption of the existence of a global $Z_2/U(1)$ symmetry, where real/complex DM representations are odd, while the Standard Model (SM) fields are even.

To obtain the theoretical predictions for DD is customary to construct an effective field theory (EFT) since the energy transfer in DD experiments is of the order of $\sim 1\rm{GeV} \ll \Lambda_\text{EW}\lesssim \Lambda_\text{BSM}$. This EFT will contain as dynamical degrees of freedom the light quark flavors ($u$, $d$ and $s$), the gluons, and the slow-varying components of the DM field as a static source. 
The full computation at loop level of the interaction from Eq.~\ref{eq:model} corresponds to the following EFT Lagrangian for the scalar DM.
\bea    
\mathcal{L}^{{\rm Scalar},q}
&=&  \sum_{q=u,d} c^q
\left( \Phi_{\rm DM}^\dagger i\stackrel{\leftrightarrow}{\partial_\mu}  \Phi_{\rm DM}\right) \bar q \gamma^\mu q + \sum_{q=u,d,s} d^q m_q \Phi_{\rm DM}^\dagger \Phi_{\rm DM}\, \bar q q+ d^g \frac{\alpha_s}{\pi}\Phi_{\rm DM}^\dagger \Phi_{\rm DM} \, G^{a\mu \nu}G^a_{\mu \nu}
\nonumber\\ 
&&
+ \sum_{q=u,d,s} \frac{g_1^q}{M_{\Phi_{\rm DM}}^2}
\Phi_{\rm DM}^\dagger (i \partial^\mu)(i \partial^\nu ) \Phi_{\rm DM}\, \mathcal{O}^{q}_{\mu \nu}
+ \frac{g_1^g}{M_{\Phi_{\rm DM}}^2}
\Phi_{\rm DM}^\dagger (i \partial^\mu)(i \partial^\nu) \Phi_{\rm DM}\, \mathcal{O}^{g}_{\mu \nu}
\ ,
\label{Scalar:leff}
\eea
where $\mathcal{O}^{q}_{\mu \nu}$ and $\mathcal{O}^{g}_{\mu \nu}$ are the twist-2 components:
\begin{equation}
\mathcal{O}^q_{\mu \nu}=\bar q \left(\dfrac{iD_\mu \gamma_\nu+iD_\nu \gamma_\mu}{2}-\frac{1}{4}g_{\mu \nu}i\slashed{D}\right)q\ ,
\quad\quad
\mathcal{O}^g_{\mu \nu}=G_\mu^{a\rho} G^a_{\nu \rho}-\frac{1}{4}g_{\mu \nu}G^a_{\rho \sigma}G^{a\rho \sigma}\,.
\end{equation}
For fermionic DM the EFT interaction is described by
\bea
\mathcal{L}^{\text{Fermion},q}
&=&
\sum_{q=u,d} c^q\, \bar\Psi_{\rm DM} \gamma_\mu \Psi_{\rm DM}\,\bar q \gamma^\mu q 
+\sum_{q=u,d,s}\tilde{c}^q\, \bar\Psi_{\rm DM} \gamma_\mu \gamma_5 \Psi_{\rm DM} \,\bar q \gamma^\mu \gamma_5 q
\nonumber\\
&&
+ \sum_{q=u,d,s} d^q\, m_q\bar \Psi_{\rm DM} \Psi_{\rm DM}\,\bar q q 
+\sum_{q=c,b,t} d_q^g\, \bar \Psi_{\rm DM}\,\Psi_{\rm DM} G^{a\mu \nu}G^a_{\mu \nu}
\nonumber\\
&&
+ \sum_{q=u,d,s} \left(g_{1}^{q} \frac{\bar \Psi_{\rm DM} i \partial^\mu \gamma^\nu \Psi_{\rm DM} \mathcal{O}^q_{\mu \nu} }{M_{\Psi_{\rm DM}}}+ g_{2}^{q} \frac{\bar \Psi_{\rm DM} \left( i  \partial^\mu \right)\left(i \partial^\nu \right) \Psi_{\rm DM} \mathcal{O}^q_{\mu \nu} }{M_{\Psi_{\rm DM}}^2}\right)
\nonumber\\
&&
+ \sum_{q=c,b,t}\left( g_{1}^{g,q} \frac{\bar \Psi_{\rm DM} i \partial^\mu \gamma^\nu \Psi_{\rm DM} \mathcal{O}^g_{\mu \nu} }{M_{\Psi_{\rm DM}}}+ g_{2}^{g,q} \frac{\bar \Psi_{\rm DM} \left( i  \partial^\mu \right)\left(i \partial^\nu \right) \Psi_{\rm DM} \mathcal{O}^g_{\mu \nu} }{M_{\Psi_{\rm DM}}^2}\right)\ .
\label{eq:Lfermion}
\eea
The complete description of the matching and the Wilson Coefficients can be found in\cite{Arcadi:2023imv}. Allowing a comparison with the current DD experiments in Fig.~\ref{fig:DD} setting the coupling to 1. 
\section{Relic Density}
A natural complementary constraint on the parameter space of the DM model comes from the DM relic density, measured with huge precision by the Planck experiment\cite{Planck:2018vyg}. On the paradigm of freeze-out, the only particle physics input accounting for the DM relic density is the thermally averaged pair annihilation cross-section through the relation
\cite{Edsjo:1997bg}:
\begin{equation}
\label{eq:relic_thermal}
\Omega_{\rm DM}h^2 \approx 8.76 \times 10^{-11}\,{\mbox{GeV}}^{-2}{\left[\int_{T_{\rm f.o.}}^{T_0} g_{*}^{1/2} \langle \sigma v \rangle_{\rm eff}\frac{dT}{M_{\rm DM}}\right]}^{-1}\,,
\end{equation}
with $g_*$ being the effective number of relativistic
degrees of freedom while, $T_{\rm f.o.} \sim \frac{M_{\rm DM}}{20}-\frac{M_{\rm DM}}{30}$ is the standard freeze-out temperature while $T_0$ is the present time temperature of the Universe. The experimentally determined value of the DM relic density \cite{Planck:2018vyg} $\Omega_{\rm DM} h^2 = 0.1199 \pm 0.0022$.

\begin{figure}[t]
    \centering
    \begin{minipage}[t]{0.40\textwidth}
        \centering
        \includegraphics[width=\textwidth]{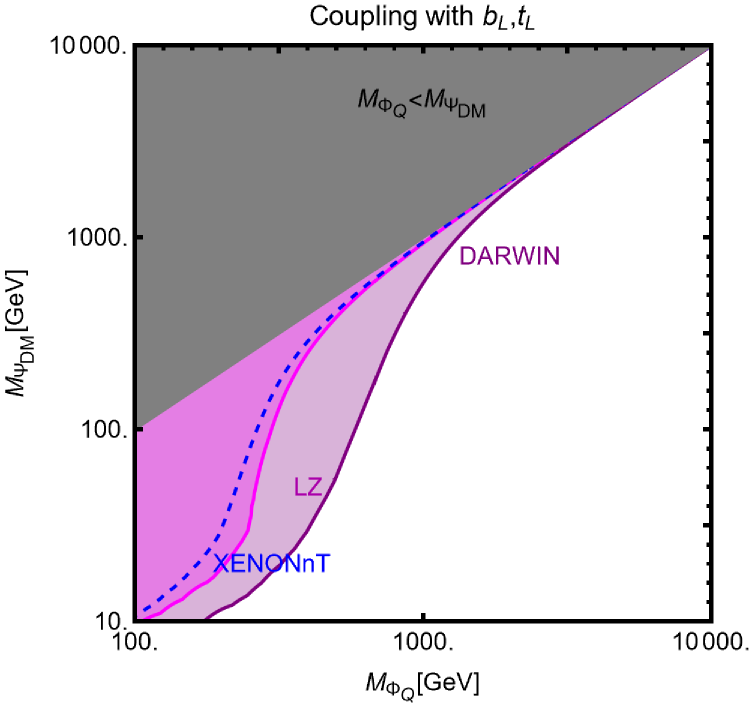}
        \caption{Direct detection prospects for Majorana DM, a singlet under the SM gauge group, involve interactions emerging at one loop. The magenta region is excluded by the latest LZ bound \protect\cite{LZ:2022lsv}. The purple region indicates the expected sensitivity of the DARWIN experiment \protect\cite{Aalbers:2016jon}. The dashed blue contour shows the bound from XENONnT \protect\cite{XENON:2023cxc}. The gray region is excluded from the analysis due to cosmological instability of the DM.}
        \label{fig:DD}
    \end{minipage}
    \begin{minipage}[t]{0.40\textwidth}
        \centering
        \includegraphics[width=\textwidth]{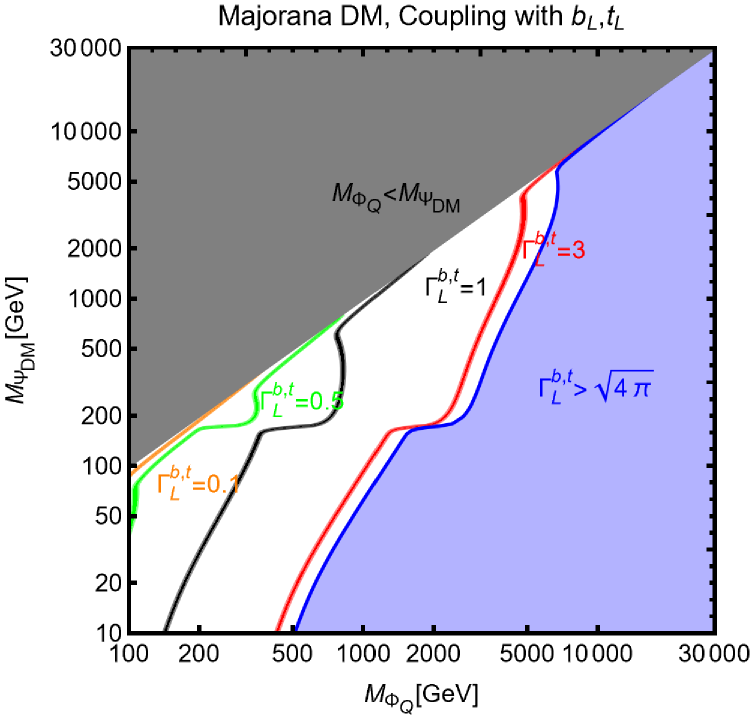}
        \caption{Isocontours of the correct DM relic density are depicted in the $(M_{\Psi_Q},M_{\Phi_{\rm DM}})$ plane for Majorana DM. Solid lines of various colors represent different coupling assignments as indicated in the panels. Gray areas indicate regions where $\Phi_{\rm DM}$ is not the DM candidate. Blue regions denote cases where the thermally favored DM annihilation cross-section can only be achieved with non-perturbative coupling values.}
        \label{fig:RD}
    \end{minipage}
\end{figure}

Despite the analytic results\cite{Arcadi:2023imv}, the Relic Density bounds presented in Fig~\ref{fig:RD} have been obtained through the package micrOMEGAs\cite{Belanger:2015nma}.
\section{Combined constrains}
\begin{figure}[t]
    \centering
    \begin{minipage}[b]{0.35\textwidth}
        \centering
        \includegraphics[width=\textwidth]{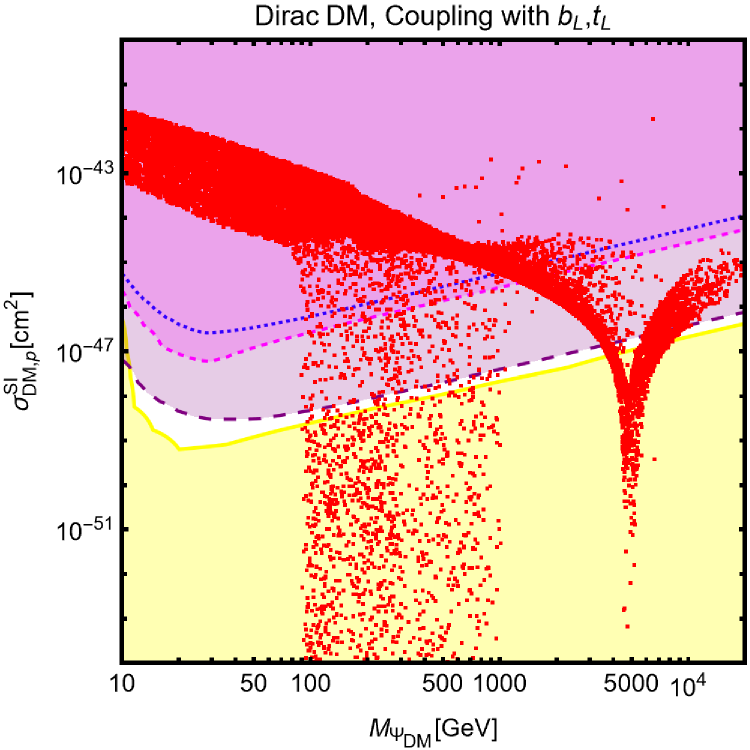}
    \end{minipage}
    \begin{minipage}[b]{0.35\textwidth}
        \centering
        \includegraphics[width=\textwidth]{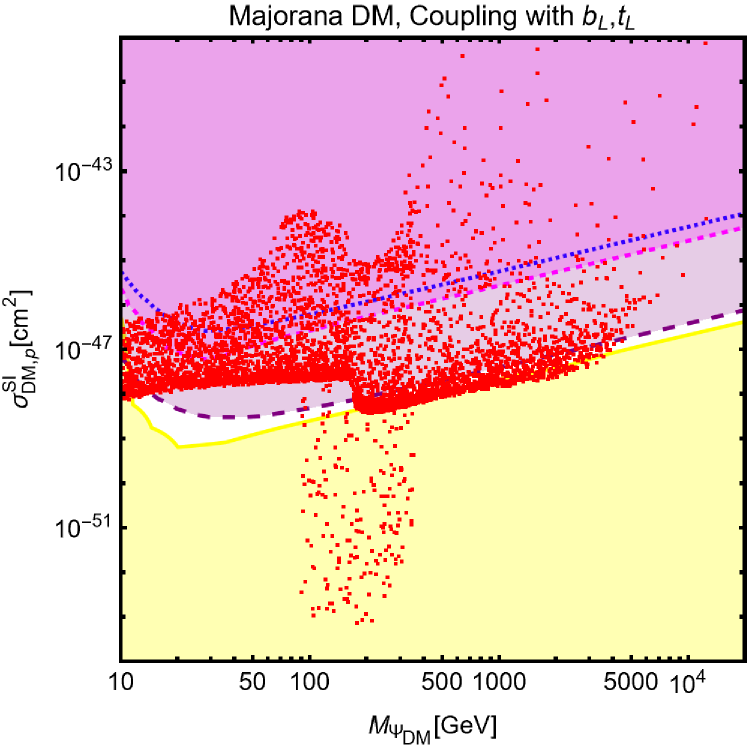}
    \end{minipage}

    \medskip 

    \begin{minipage}[b]{0.35\textwidth}
        \centering
        \includegraphics[width=\textwidth]{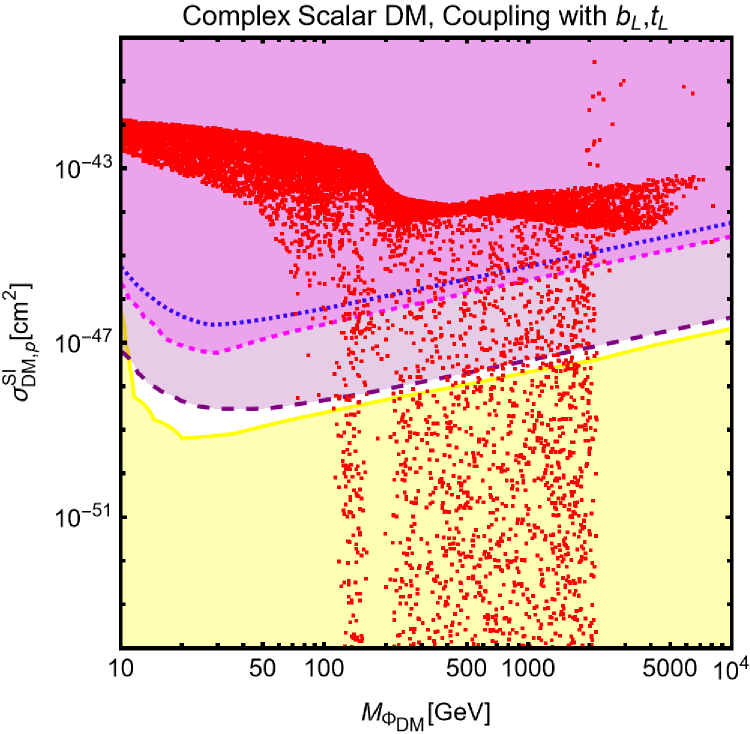}
    \end{minipage}
    \begin{minipage}[b]{0.35\textwidth}
        \centering
        \includegraphics[width=\textwidth]{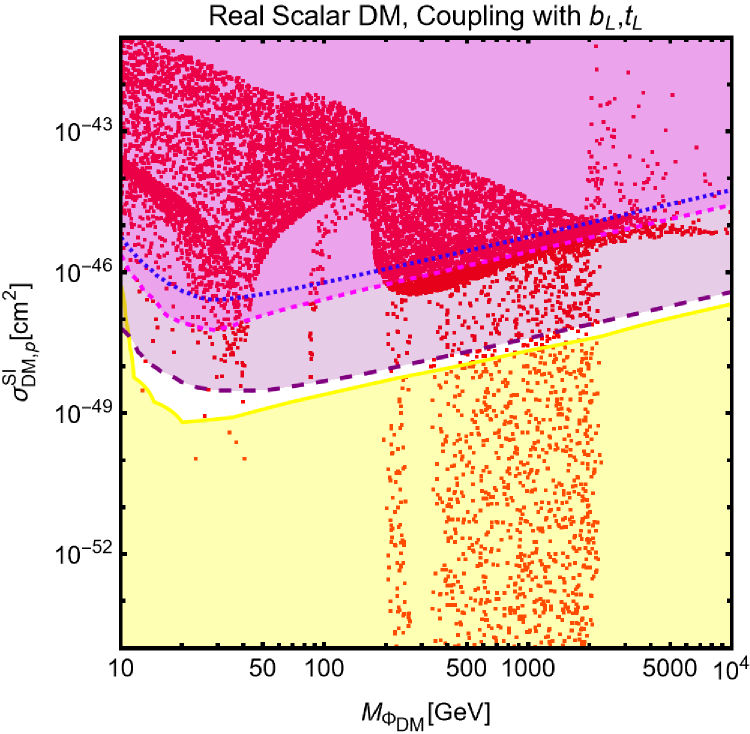}
    \end{minipage}
    \caption{Parameter spaces satisfying the correct relic density are plotted in the $(M_{\Phi_{\rm DM}},\sigma_{\Phi_{\rm DM},p}^{\rm SI})$ two-dimensional plane. Exclusions by LZ (magenta region), XENONnT (region above the blue dashed line), the projected sensitivity by DARWIN (purple region), and the region corresponding to the neutrino floor (yellow region) are highlighted.}
    \label{fig:3gen}
\end{figure}

We combine more systematically the constraints obtained by DM DD as well as the relic density, considering a minimal scenario where DM candidate only couples with a single quark species, mediated by a single BSM particle. To analyze the entire phase space, we scan over the following parameter range:
\bea
\label{eq:min_scan}
    & M_{\Phi_{\rm DM},\Psi_{\rm DM}}\in \left[10,10^5\right]\,\mbox{GeV},\qquad
     M_{\Phi_f,\Psi_f} \in \left[100,10^5\right]\,\mbox{GeV},\qquad
     \Gamma_{L,R}^{f}\in \left[10^{-3},\sqrt{4\pi}\right],
\eea
and retained only the model points, i.e. assignation for the set $(M_{\Psi_{\rm DM},\Phi_{\rm DM}},M_{\Phi_{f},\Psi_{f}},\Gamma_{L,R}^f)$, giving the correct DM relic density. We confront the result with the current DD bounds at Fig.~\ref{fig:3gen}.
\section{Conclusion}
 In this talk, we presented updated bounds on t-channel Dark Matter models, as given in ref\cite{Arcadi:2023imv} which also showed the complete matching at the one-loop level. We compare the results with current Direct Detection (DD) and relic density constraints, revealing strong limitations on both the Complex and Dirac cases, except for the finely tuned coannihilation region. The Real DM scenario presents weaker Direct Detection constraints but, owing to a significantly suppressed annihilation cross-section, still imposes notable restrictions on the candidate. Finally, among the scenarios considered in this study, Majorana DM emerges as the most favored one, being the only one allowing for viable masses of approximately 100 $\rm{GeV}$ or lower

\section*{Acknowledgments}
D.C.A. acknowledges funding from the Spanish MCIN/AEI/10.13039/501100011033 through grant
CNS2022-135262 funded by the “European Union NextGenerationEU/PRTR”.
\section*{References}

\label{Bibliography}

\bibliography{Bibl} 

\end{document}